\documentclass[12pt,a4paper]{article}
\usepackage{amsmath,amsfonts,amsthm,graphicx,lscape}
\usepackage[dvips]{psfrag}
\usepackage{cite}

\usepackage[normalem]{ulem}

\usepackage{textcomp}
\usepackage{amsmath,amsthm,amssymb,mathrsfs,cases}
\usepackage{amsfonts,graphicx,lscape}

\newcommand{\sca}[2]{\langle #1, #2 \rangle}

\usepackage{accents}
\makeatletter
\def\widebar{\accentset{{\cc@style\underline{\mskip10mu}}}}
\makeatother

\numberwithin{equation}{section} 
\newtheorem{theorem}{Theorem}[section]
\newtheorem{proposition}[theorem]{Proposition}

\def\beqa{\begin{eqnarray}}
\def\enqa{\end{eqnarray}}
\def\beq{\begin{equation}}
\def\enq{\end{equation}}

\begin{document}
\title{
Integrable semi-discretizations 
of the Davey--Stewartson system 
and a \mbox{$(2+1)$}-dimensional
Yajima--Oikawa system. I
}
\author{Takayuki \textsc{Tsuchida}
}
\maketitle
\begin{abstract} 
%
The integrable Davey--Stewartson system 
is a linear combination 
of 
the two elementary flows that commute:  
$\mathrm{i} q_{t_1} + q_{xx} + 2q\partial_y^{-1}\partial_x (|q|^2) =0$
and $\mathrm{i} q_{t_2} + q_{yy} + 2q\partial_x^{-1}\partial_y (|q|^2) =0$. 
In the literature, 
each 
elementary Davey--Stewartson 
flow is 
often called 
the Fokas system 
%
because it was studied by Fokas 
in the early 1990s. 
In fact, the integrability of the Davey--Stewartson system 
dates back to the work of Ablowitz and Haberman in 1975; 
the elementary Davey--Stewartson flows, 
as well as 
another integrable 
\mbox{$(2+1)$}-dimensional 
nonlinear Schr\"odinger equation 
$\mathrm{i} q_{t} + q_{xy} + 2 q\partial_y^{-1}\partial_x (|q|^2) =0$
proposed by Calogero and Degasperis in 1976, 
appeared explicitly in Zakharov's article 
published 
in 1980. 
By applying a linear change of the independent variables, 
an elementary Davey--Stewartson flow 
can be 
identified 
with 
a 
\mbox{$(2+1)$}-dimensional generalization 
of the integrable long wave--short wave interaction
model, 
called the Yajima--Oikawa system: 
$\mathrm{i} q_{t} + q_{xx} + u q=0$,  
$u_t + c u_y = 2(|q|^2)_x$. 
In this paper, we propose 
a new integrable 
semi-discretization 
(discretization of one of the two spatial variables, say 
$x$)  
of the Davey--Stewartson system by constructing 
its Lax-pair representation; the two elementary flows 
in the semi-discrete case indeed commute. 
By applying a 
linear change of the continuous independent variables 
to an elementary flow, 
we also obtain an integrable 
semi-discretization of the \mbox{$(2+1)$}-dimensional 
Yajima--Oikawa system.

\end{abstract}
%
%
\newpage
\noindent
\tableofcontents

\newpage
\section{Introduction}
The 
Davey--Stewartson system~\cite{DS74}
(also
known as the Benney--Roskes system~\cite{Benney69})
is a \mbox{$(2+1)$}-dimensional generalization of 
the nonlinear Schr\"odinger equation, 
which 
can be classified into three different types~\cite{BPS1993}; 
two of 
the three types 
can be written as~\cite{Nizh82}
\begin{subequations}
\label{continuousDS}
\begin{equation}
\mathrm{i} q_{t} + a \left( q_{xx} + 2F q \right) + b \left( q_{yy} + 2G q \right) =0. 
\end{equation}
Here, $a$ and $b$ are real constants, the subscripts denote the partial differentiation 
and $F$ (in the case of \mbox{$a \neq 0$}) and $G$ (in the case of \mbox{$b \neq 0$}) are nonlocal 
real-valued 
potentials defined as 
\begin{equation}
 F_y := (|q|^2)_x, \hspace{5mm} G_x := (|q|^2)_y.
\end{equation}
\end{subequations}
The Davey--Stewartson system
of the remaining type 
corresponds to the ``complex-valued'' case and 
can be written as 
\begin{equation}
\begin{cases}
\mathrm{i} q_{t} + a \left[ \left( \partial_x + \mathrm{i} \partial_y \right)^2 q + 2F q \right] 
	+ a^\ast \left[ \left( \partial_x - \mathrm{i} \partial_y \right)^2 q + 2F^\ast q \right] =0, 
\nonumber
\\[2pt]
\left( \partial_x - \mathrm{i} \partial_y \right) F = \sigma \left( \partial_x + \mathrm{i} \partial_y \right) (|q|^2).
\nonumber
\end{cases}
\end{equation}
Here, \mbox{$a \in \mathbb{C}$},  \mbox{$\sigma \in \mathbb{R}$}, 
the nonlocal 
potential $F$ is complex-valued 
and the asterisk denotes the complex conjugate, 
but this type is outside the scope of this paper. 

The integrability of 
the Davey--Stewartson system 
can be traced back to the work of Ablowitz and Haberman in 1975~\cite{Hab75} 
(also see~\cite{Morris77,Anker,Ab78,Cornille}), 
who gave its 
Lax-pair representation~\cite{Lax} up to a coordinate transformation; 
note that the Lax-pair representation 
in \mbox{$2+1$} dimensions 
expressed in operator form 
is often referred to as 
the Manakov triad 
representation~\cite{Manakov_triad}. 

The Davey--Stewartson system (\ref{continuousDS})
is a linear combination 
of 
the two elementary flows: 
\begin{equation}
\mathrm{i} q_{t_1} + q_{xx} + 2F q =0, \hspace{5mm} F_y= (|q|^2)_x,
\label{element1}
\end{equation}
and 
\begin{equation}
\mathrm{i} q_{t_2} + q_{yy} + 2G q =0, \hspace{5mm} G_x = (|q|^2)_y.
\label{element2}
\end{equation} 
The elementary Davey--Stewartson 
flow (\ref{element1}) 
(or (\ref{element2})) 
is 
often referred to as 
the Fokas system 
because it appeared in his paper~\cite{Fokas94} published in 1994. 
Note, however, that 
the elementary Davey--Stewartson 
flow (\ref{element1}), 
as well as 
another integrable 
\mbox{$(2+1)$}-dimensional 
generalization of the 
nonlinear Schr\"odinger equation 
\mbox{$\mathrm{i} q_{t} + q_{xy} + 2 F q=0$}, \mbox{$F_y= (|q|^2)_x$} 
originally 
proposed by Calogero and Degasperis~\cite{Calo76} in 1976, 
appeared 
explicitly 
in Zakharov's article~\cite{Zakh}
published 
in 1980. 
In addition, 
a linear change of the independent variables (see, {\em e.g.}, page 135 of 
\cite{Kono92}): 
\begin{equation}
\widetilde{t} = t_1 + y, \hspace{5mm} \widetilde{x}=x, \hspace{5mm} \widetilde{y}= c y, 
\label{Galilean-like}
\end{equation}
with an arbitrary real constant $c$,   
converts the elementary Davey--Stewartson flow (\ref{element1}) 
to a \mbox{$(2+1)$}-dimensional generalization of an integrable 
long wave--short wave interaction model
(known as the Yajima--Oikawa system~\cite{YO76}): 
\begin{equation}
\mathrm{i} q_{t} + q_{xx} + u q =0, \hspace{5mm} u_t + c u_y= 2 (|q|^2)_x,
\label{2DYO}
\end{equation}
which was proposed by Mel'nikov~\cite{Mel83} in 1983. 
Here, \mbox{$u:=2F$} and the tilde is omitted for notational brevity. 
The \mbox{$(2+1)$}-dimensional Yajima--Oikawa system 
(\ref{2DYO}) with \mbox{$c=1$} is often referred to as 
the Maccari system~\cite{Maccari1}. 

From (\ref{element1}) and (\ref{element2}), we 
obtain the relation \mbox{$F_{y t_2}=\mathrm{i} \left( q_y q^\ast - q q^\ast_y \right)_{y x}$}, 
which implies that 
\begin{equation}
F_{t_2} = \mathrm{i} \left( q_y q^\ast - q q^\ast_y \right)_x + \xi, \hspace{5mm} \xi_y=0. 
\label{F_{t_2}}
\end{equation}
%
Similarly, 
we also 
obtain the relation 
\mbox{$G_{x t_1}=\mathrm{i} \left( q_x q^\ast - q q^\ast_x \right)_{x y}$}, 
which implies that 
\begin{equation}
G_{t_1} = \mathrm{i} \left( q_x q^\ast - q q^\ast_x \right)_y + \eta, \hspace{5mm} \eta_x=0. 
\label{G_{t_1}}
\end{equation}
With the aid of (\ref{F_{t_2}}) and (\ref{G_{t_1}}), we can show that 
the two elementary 
flows (\ref{element1}) and (\ref{element2}) commute~\cite{Fokas94,Kaji90}, i.e., 
\mbox{$q_{t_1 t_2} = q_{t_2 t_1}$}
if and only if \mbox{$\xi=\eta$}; 
in this case, we can set without loss of generality 
\mbox{$\xi=\eta=0$}
by a change of variables
\mbox{$q= \widetilde{q} \exp \left( 2 \mathrm{i} \int \int \xi \, \mathrm{d} t_1 \mathrm{d} t_2 \right)$}.
In short, the commutativity of the elementary flows\footnote{The author is indebted to 
Dr.~Masato Hisakado and Professor Aristophanes Dimakis
for the proof of commutativity.} is guaranteed only if the 
``constants'' of integration in the nonlocal 
potentials $F$ and $G$ are chosen appropriately. 
More specifically, the two elementary flows 
commute if and only if the $y$-independent value of 
\mbox{$F_{t_2} - \mathrm{i} \left( q_y q^\ast - q q^\ast_y \right)_x $} 
evaluated at any fixed value of $y$ 
is equal to the $x$-independent value of 
\mbox{$G_{t_1} - \mathrm{i} \left( q_x q^\ast - q q^\ast_x \right)_y$} 
evaluated at any fixed value of $x$, 
which is thus \mbox{$(x,y)$}-independent and can be set equal to zero 
by redefining the dependent variable $q$. 

In this and the next paper, we propose 
integrable semi-discretizations 
of the Davey--Stewartson system (\ref{continuousDS})  
and the \mbox{$(2+1)$}-dimensional Yajima--Oikawa system (\ref{2DYO}) 
by constructing their Lax-pair representations. 
We consider the discretization of only one of the spatial variables $x$ and $y$, 
which 
is referred to as 
a ``semi-discretization'' in this and the next paper. 
This is in 
contrast to our previous work on an integrable discretization 
of the Davey--Stewartson system~\cite{TD11} (also see some preceding 
studies in~\cite{Hu06,Hu07}) wherein both spatial variables are discretized. 
%
The semi-discrete Davey--Stewartson system 
proposed in this paper is not obtainable 
by taking a continuous limit of one spatial variable 
in the discrete Davey--Stewartson system proposed in our previous paper~\cite{TD11}. 

This paper is organized as follows. 
In section 2, we propose 
an 
integrable semi-discretization 
for each 
of the 
elementary Davey--Stewartson flows 
(\ref{element1}) and (\ref{element2}) by constructing 
its Lax-pair representation. 
In section 3, 
we 
demonstrate that 
these 
semi-discretizations 
commute under a suitable choice of 
the ``constants'' of integration as in the continuous case. 
Then, 
we consider 
a linear combination of the 
semi-discrete elementary Davey--Stewartson flows 
to obtain 
an integrable semi-discretization of the full Davey--Stewartson system (\ref{continuousDS}). 
Moreover, by applying 
a linear change of the independent variables like 
(\ref{Galilean-like}) to 
one of the semi-discrete elementary Davey--Stewartson flows, 
we also obtain an integrable semi-discretization of 
the \mbox{$(2+1)$}-dimensional Yajima--Oikawa system (\ref{2DYO}) 
(see 
a 
relevant 
semi-discrete system 
in~\cite{Yu15}).  
Section 4 is devoted to concluding remarks.

\section{
Integrable semi-discretizations of 
the 
two 
elementary Davey--Stewartson flows}

\subsection{Semi-discrete linear 
problem} 

The continuous Davey--Stewartson system (\ref{continuousDS}) 
is obtained as the compatibility conditions of the overdetermined linear 
systems for $\psi$ and $\phi$~\cite{Hab75,Nizh82}:
\begin{subnumcases}{\label{clinear_s}}
\psi_{y} = q \phi, 
\label{clinear_s1}
\\
\phi_{x} = -q^\ast \psi, 
\label{clinear_s2}
\end{subnumcases}
and 
\begin{subnumcases}{\label{clinear_t}}
\mathrm{i} \psi_{t} = -a \psi_{xx} - 2a F \psi + b q \phi_y - b q_y \phi, 
\label{clinear_t1}
\\
\mathrm{i} \phi_{t} = a q^\ast \psi_x - a q_x^\ast \psi + b \phi_{yy} + 2 b G \phi. 
\label{clinear_t2}
\end{subnumcases}
This Lax-pair representation for 
the Davey--Stewartson system 
can be straightforwardly 
generalized 
to the case of 
vector- or 
matrix-valued dependent variables~\cite{Kono92,Calogero91,
Fordy87,March}. 

As a semi-discrete analog of 
the spatial part of the Lax-pair representation for 
(the vector generalization of) the Davey--Stewartson system, 
we 
consider the 
following 
linear problem in 
two spatial dimensions: 
\begin{subnumcases}{\label{sdlinear}}
\psi_{n,y} = q_n \left( \gamma \phi_{n} + \delta \phi_{n+1} \right), 
\label{sdlinear1}
\\[2pt]
\phi_{n+1}-\phi_{n} = r_n \psi_n. 
\label{sdlinear2}
\end{subnumcases}
Here, $n$ is a discrete spatial variable and $y$ is a continuous spatial variable; 
the subscript $y$ denotes the differentiation with respect to $y$. 
The 
constants $\gamma$ and $\delta$ 
should 
satisfy 
the condition \mbox{$\gamma+\delta \neq 0$}; 
\mbox{$\gamma+\delta$} can be fixed at any nonzero value 
by rescaling the dependent variable $q_n$, 
so only the ratio $\gamma/\delta$ is essential. 
The dependent variables $q_n$ and $r_n$
are 
$M$-component 
row 
and 
column 
vectors, respectively; 
a scalar component $\psi_n$ 
and an $M$-component 
column vector 
$\phi_n$ 
comprise 
the linear wavefunction. 
We do not distinguish between 
the left scalar multiplication and the right scalar multiplication, 
so, for example, \mbox{$q_n \gamma \phi_{n} = \gamma q_n \phi_{n} = q_n \phi_{n} \gamma$} 
in (\ref{sdlinear1}). 
%
Note that using the second equation (\ref{sdlinear2}), the first equation (\ref{sdlinear1}) can be 
rewritten as 
\begin{equation}
\nonumber
\psi_{n,y}= \left( \gamma + \delta \right) q_n \phi_n + \delta q_n r_n \psi_n, 
\end{equation}
where $\psi_n$, $q_n \phi_n$ and $q_n r_n$ are 
scalar functions.
%


\subsection{Semi-discretization 
of the elementary Davey--Stewartson flow (\ref{element1})}

One possible
choice of time evolution of 
the linear wavefunction
is 
\begin{subnumcases}{\label{sd_time1}}
\mathrm{i} \psi_{n,t_1} = \alpha u_n^\gamma \psi_{n-1} + \beta u_{n+1}^\delta \psi_{n+1} 
	+ w_n \psi_{n}, 
\label{sd_time1_1}
\\[2pt]
\mathrm{i} \phi_{n,t_1} =  -\alpha u_n^\gamma r_n \psi_{n-1}  + \beta u_{n}^\delta r_{n-1} \psi_{n},
\label{sd_time1_2}
\end{subnumcases}
%
where $\alpha$ and $\beta$ are 
constants
satisfying the condition
\mbox{$(\alpha \gamma, \beta \delta) \neq (0,0)$}, 
and $u_n$ and 
$w_n$ 
are 
scalar functions. 


\begin{proposition}
\label{prop2.1}
The compatibility 
conditions of the 
overdetermined 
linear 
systems 
\mbox{$(\ref{sdlinear})$} and \mbox{$(\ref{sd_time1})$} 
for $\psi_n$ and $\phi_n$ 
are equivalent to the system of 
differential-difference equations: 
\begin{equation} 
\label{first_flow}
\left\{ 
\begin{split}
& \mathrm{i} q_{n,t_1} - \alpha u_n^\gamma q_{n-1} - \beta u_{n+1}^\delta q_{n+1} - w_n q_n =0, 
\\[2pt]
& \mathrm{i} r_{n,t_1} + \beta u_{n}^\delta r_{n-1} + \alpha u_{n+1}^\gamma r_{n+1} +w_n r_n  =0, 
\\[2pt]
& u_{n,y} = u_n \left( q_{n-1} r_{n-1} - q_n r_n \right), 
\\[2pt]
& w_{n,y} = \alpha \delta \left( u_n^{\gamma} q_{n-1} r_{n} 
	- u_{n+1}^{\gamma} q_n r_{n+1} \right) 
	+ \beta \gamma \left( u_n^{\delta} q_{n} r_{n-1} 
	- u_{n+1}^{\delta} q_{n+1} r_{n} \right).
\end{split} 
\right. 
\end{equation}
\end{proposition}
This proposition can be proved by a direct calculation. 
Indeed, using (\ref{sdlinear}) and (\ref{sd_time1}),  we have 
\begin{align}
0 & = \mathrm{i} \psi_{n,y t_1} - \mathrm{i} \psi_{n,t_1 y} 
\nonumber \\
& = \left( \mathrm{i} q_{n,t_1} - \alpha u_n^\gamma q_{n-1} - \beta u_{n+1}^\delta q_{n+1} - w_n q_n \right) 
	\left( \gamma \phi_n + \delta \phi_{n+1} \right)
\nonumber \\
& \hphantom{=} \; \, \mbox{} 
	+ \alpha \gamma u_n^{\gamma-1} \left( - u_{n,y} + u_n q_{n-1} r_{n-1} - u_n q_n r_n \right) \psi_{n-1}
\nonumber \\
& \hphantom{=} \; \, \mbox{}
	+ \left( - w_{n,y} + \alpha \delta u_n^{\gamma} q_{n-1} r_{n} - \alpha \delta u_{n+1}^{\gamma} q_n r_{n+1} 
	+ \beta \gamma u_n^{\delta} q_{n} r_{n-1} - \beta \gamma u_{n+1}^{\delta} q_{n+1} r_{n} \right) \psi_n 
\nonumber \\
& \hphantom{=} \; \, \mbox{}
	+ \beta \delta u_{n+1}^{\delta -1} \left( - u_{n+1,y} + u_{n+1} q_{n} r_{n} - u_{n+1} q_{n+1} r_{n+1} \right) \psi_{n+1}, 
\nonumber 
\end{align}
and 
\begin{align}
0 &= \mathrm{i} \left( r_{n} \psi_{n} + \phi_{n} - \phi_{n+1} \right)_{t_1} 
\nonumber \\
& = \left( \mathrm{i} r_{n,t_1} + \beta u_{n}^\delta r_{n-1} + \alpha u_{n+1}^\gamma r_{n+1} +w_n r_n \right) \psi_{n},
\nonumber 
\end{align}
which imply (\ref{first_flow}) for 
generic 
$\psi_n$ and $\phi_n$. 

Under the parametric conditions
\begin{equation}
\beta=\alpha^\ast, \hspace{5mm} \gamma=\delta \in \mathbb{R}, 
\nonumber
\end{equation}
the system (\ref{first_flow}) 
admits 
the Hermitian conjugation reduction: 
\begin{equation}
r_n =  - \varDelta q_n^\dagger, \hspace{5mm} u_n^\ast=u_n, \hspace{5mm} w_n^\ast = w_n,  
\nonumber
\end{equation}
where 
$\varDelta$ is 
an arbitrary real constant 
that will be interpreted as a lattice parameter 
and the dagger denotes the 
Hermitian conjugation. 
In particular, 
if \mbox{$\alpha=\beta=\gamma=\delta=1$}, 
this reduction simplifies 
(\ref{first_flow}) to 
\begin{equation} 
\label{reduced_first_flow2}
\left\{ 
\begin{split}
& \mathrm{i} q_{n,t_1} 
= u_{n+1} q_{n+1} + w_n q_n + u_n q_{n-1} 
, 
\\[2pt]
& u_{n,y} = \varDelta u_n \left( \sca{q_n}{q_n^\ast} 
- \sca{q_{n-1}}{q_{n-1}^\ast} \right) 
, 
\\[2pt]
& w_{n,y} = \varDelta u_{n+1} \left( \sca{q_n}{q_{n+1}^\ast} + \sca{q_{n+1}}{q_n^\ast} \right)
 - \varDelta u_n \left( \sca{q_{n-1}}{q_n^\ast} + \sca{q_n}{q_{n-1}^\ast} \right),
\end{split} 
\right. 
\end{equation}
where \mbox{$q_n \in \mathbb{C}^M$}, 
\mbox{$u_n, w_n \in \mathbb{R}$} 
and \mbox{$\sca{\,\cdot\,}{\,\cdot\,}$} stands for 
the standard 
scalar product. 


In the case where \mbox{$M=1$}, i.e., $q_n$ is a scalar, 
(\ref{reduced_first_flow2}) provides an integrable semi-discretization 
of the elementary Davey--Stewartson flow (\ref{element1}). 
Indeed, 
by setting 
\begin{equation}
\nonumber
q_n = q(n \Delta, y, t_1), \hspace{5mm} u_n = \frac{1}{\Delta^2} + \frac{1}{2} u(n \Delta, y, t_1), 
\hspace{5mm} w_n = -\frac{2}{\Delta^2} + w(n \Delta, y, t_1),
\end{equation}
and taking the continuous limit \mbox{$\Delta \to 0$}, 
(\ref{reduced_first_flow2})
with 
scalar $q_n$
reduces to
\begin{equation} 
\label{reduced_first_flow4}
\left\{ 
\begin{split}
& \mathrm{i} q_{t_1} 
= q_{xx} + \left( u+w \right) q, 
\\[2pt]
& u_{y} =  w_{y} = 2 \left( \left| q \right|^2 \right)_x, 
\end{split} 
\right. 
\end{equation}
where \mbox{$x:= n \Delta$}. 
Thus, (\ref{reduced_first_flow4}) for 
the pair of dependent variables 
\mbox{$(q, u+w)$}
can be 
identified with the elementary Davey--Stewartson flow (\ref{element1}), 
up to a rescaling of variables. 

Actually,  the 
differentiation operator $\partial_{t_1}$ in (\ref{sd_time1})
can be represented as 
a linear combination 
of two elementary differentiation operators 
as 
\begin{equation}
\nonumber
\mathrm{i} \partial_{t_1} = \alpha \partial_{t_\alpha} + \beta \partial_{t_\beta}. 
\end{equation}
Here, the 
two 
operators 
$\partial_{t_\alpha}$ and $\partial_{t_\beta}$ correspond to 
the case \mbox{$\alpha=1$}, \mbox{$\beta=0$} 
and the case \mbox{$\alpha=0$}, \mbox{$\beta=1$}, respectively, i.e.,  
\begin{subnumcases}{\label{sd_time_a}}
\psi_{n,t_\alpha} = u_n^\gamma \psi_{n-1} + w^{(\alpha)}_n \psi_{n}, 
\label{sd_time_a_1}
\\[2pt]
\phi_{n,t_\alpha} =  -u_n^\gamma r_n \psi_{n-1},
\label{sd_time_a_2}
\end{subnumcases}
and 
\begin{subnumcases}{\label{sd_time_b}}
\psi_{n,t_\beta} = u_{n+1}^\delta \psi_{n+1} + w^{(\beta)}_n \psi_{n}, 
\label{sd_time_b_1}
\\[2pt]
\phi_{n,t_\beta} =  u_{n}^\delta r_{n-1} \psi_{n},
\label{sd_time_b_2}
\end{subnumcases}
where \mbox{$w_n = \alpha w^{(\alpha)}_n + \beta w^{(\beta)}_n $}. 

As special cases of Proposition~\ref{prop2.1}, we obtain the 
following two propositions. 
\begin{proposition}
\label{prop2.2}
The compatibility 
conditions of the 
overdetermined 
linear 
systems 
\mbox{$(\ref{sdlinear})$} and \mbox{$(\ref{sd_time_a})$} 
for $\psi_n$ and $\phi_n$ 
are equivalent to the system of 
differential-difference equations: 
\begin{equation} 
\label{first_flow_a}
\left\{ 
\begin{split}
& q_{n,t_\alpha} = u_n^\gamma q_{n-1} + w_n^{(\alpha)} q_n, 
\\[2pt]
& r_{n,t_\alpha} = -u_{n+1}^\gamma r_{n+1} - w_n^{(\alpha)} r_n, 
\\[2pt]
& u_{n,y} = u_n \left( q_{n-1} r_{n-1} - q_n r_n \right), 
\\[2pt]
& w^{(\alpha)}_{n,y} = \delta \left( u_n^{\gamma} q_{n-1} r_{n} 
	- u_{n+1}^{\gamma} q_n r_{n+1} \right). 
\end{split} 
\right. 
\end{equation}
\end{proposition}

\begin{proposition}
\label{prop2.3}
The compatibility 
conditions of the 
overdetermined 
linear 
systems 
\mbox{$(\ref{sdlinear})$} and \mbox{$(\ref{sd_time_b})$} 
for $\psi_n$ and $\phi_n$ 
are equivalent to the system of 
differential-difference equations: 
\begin{equation} 
\label{first_flow_b}
\left\{ 
\begin{split}
& q_{n,t_\beta} = u_{n+1}^\delta q_{n+1} + w_n^{(\beta)} q_n, 
\\[2pt]
& r_{n,t_\beta} = - u_{n}^\delta r_{n-1} - w_n^{(\beta)} r_n, 
\\[2pt]
& u_{n,y} = u_n \left( q_{n-1} r_{n-1} - q_n r_n \right), 
\\[2pt]
& w^{(\beta)}_{n,y} = \gamma \left( u_n^{\delta} q_{n} r_{n-1} 
	- u_{n+1}^{\delta} q_{n+1} r_{n} \right).
\end{split} 
\right. 
\end{equation}
\end{proposition}

The system (\ref{first_flow_a}) implies the relation: 
\begin{equation}
\nonumber
\left( \log u_n^\delta \right)_{y t_\alpha}= w^{(\alpha)}_{n-1,y} - w^{(\alpha)}_{n,y}. 
\end{equation}
Thus, by assuming that \mbox{$\left( \log u_n^\delta \right)_{t_\alpha} - w^{(\alpha)}_{n-1} + w^{(\alpha)}_{n} = 0$} 
at some
value of $y$ (say, $-\infty$, $0$, or $+\infty$), 
we obtain 
\begin{equation}
\label{u_n_t_alpha}
\left( \log u_n^\delta \right)_{t_\alpha}= w^{(\alpha)}_{n-1} - w^{(\alpha)}_{n}. 
\end{equation}
The system (\ref{first_flow_b}) implies the relation: 
\begin{equation}
\nonumber
\left( \log u_n^\gamma \right)_{y t_\beta}= w^{(\beta)}_{n,y} - w^{(\beta)}_{n-1,y}. 
\end{equation}
Thus, by assuming that \mbox{$\left( \log u_n^\gamma \right)_{t_\beta} - w^{(\beta)}_{n} + w^{(\beta)}_{n-1} = 0$} 
at some 
value of $y$ (say, $-\infty$, $0$, or $+\infty$), we obtain
\begin{equation}
\label{t_n_t_beta}
\left( \log u_n^\gamma \right)_{t_\beta}= w^{(\beta)}_{n} - w^{(\beta)}_{n-1}. 
\end{equation}
From (\ref{first_flow_a})--(\ref{t_n_t_beta}), 
we obtain 
\begin{equation}
\nonumber
\left( w_n^{(\alpha)} \right)_{y t_\beta}= \frac{\delta}{\gamma + \delta} \left( u_{n+1}^{\gamma + \delta} 
	- u_n^{\gamma + \delta} \right)_{y}, 
\end{equation}
and 
\begin{equation}
\nonumber
\left( w_n^{(\beta)} \right)_{y t_\alpha}= \frac{\gamma}{\gamma + \delta} \left( u_{n}^{\gamma + \delta} 
	- u_{n+1}^{\gamma + \delta} \right)_{y}, 
\end{equation}
which imply 
\begin{equation}
\label{w_n_t_beta}
\left( w_{n}^{(\alpha)} \right)_{t_\beta} = \frac{\delta}{\gamma + \delta} \left( u_{n+1}^{\gamma + \delta} 
	- u_n^{\gamma + \delta} \right) + J_n, 
\end{equation}
and 
\begin{equation}
\label{w_n_t_alpha}
\left( w_{n}^{(\beta)} \right)_{t_\alpha} = \frac{\gamma}{\gamma + \delta} \left( u_{n}^{\gamma + \delta} 
	- u_{n+1}^{\gamma + \delta} \right) + K_n, 
\end{equation}
respectively. Here, the ``constants'' of integration $J_n$ and $K_n$ are $y$-independent scalars. 
By a direct calculation, we 
arrive at the necessary and sufficient condition for 
the commutativity of the 
two 
operators 
$\partial_{t_\alpha}$ and $\partial_{t_\beta}$. 
\begin{proposition}
\label{}
Equations 
\mbox{$(\ref{first_flow_a})$}--\mbox{$(\ref{w_n_t_alpha})$}
imply 
that 
the two differentiation operators $\partial_{t_\alpha}$ and $\partial_{t_\beta}$ commute, i.e., 
\begin{equation}
\nonumber 
q_{n, t_\alpha t_\beta} = q_{n, t_\beta t_\alpha} \;\, \mathrm{and} \;\, r_{n, t_\alpha t_\beta} = r_{n, t_\beta t_\alpha}, 
\end{equation}
if and only if  the ``constants'' of integration $J_n$ and $K_n$ 
satisfy the condition \mbox{$J_n=K_n$}. 
\end{proposition}

\subsection{Semi-discretization
of the elementary Davey--Stewartson flow (\ref{element2})}

Another possible choice of time evolution of the linear wavefunction is 
given by 
\begin{subnumcases}{\label{sd_time2}}
k \psi_{n,t_{2}} = \left( \gamma + \delta \right) q_n \phi_{n,y} - \left( \gamma + \delta \right) 
	\left( q_{n,y} - \delta q_n r_n q_n \right) \phi_n
\nonumber \\ \hspace{15mm} \mbox{}
	- \delta \left[ \left( q_{n,y} r_n - q_n r_{n,y} \right) 
	+ \left( \gamma - \delta \right) \left( q_n r_n \right)^2 \right] \psi_{n}, 
\label{sd_time2_1}
\\[4pt]
k \phi_{n,t_{2}} =  \phi_{n,yy}  + 
	\left( \gamma + \delta \right) v_{n} \phi_{n}, 
\label{sd_time2_2}
\end{subnumcases}
where $k$ is an arbitrary (but nonzero) constant and 
$v_n$ is an \mbox{$M \times M$} square matrix. 

\begin{proposition}
\label{}
The compatibility 
conditions of the 
overdetermined 
linear 
systems 
\mbox{$(\ref{sdlinear})$} and \mbox{$(\ref{sd_time2})$} 
for $\psi_n$ and $\phi_n$ 
are equivalent to the system of 
differential-difference equations: 
\begin{equation} 
\label{second_flow}
\left\{ 
\begin{split}
& k q_{n,t_{2}} + q_{n,yy} + q_n \left( \gamma v_n + \delta v_{n+1} \right) + \gamma \delta \left( q_n r_n \right)^2 q_n =0, 
\\[2pt]
& k r_{n,t_{2}} - r_{n,yy} - \left( \delta v_n + \gamma v_{n+1} \right) r_n - \gamma \delta r_n \left( q_n r_n \right)^2 =0, 
\\[2pt]
& v_{n+1}-v_{n} = - 2 \left( r_n q_n \right)_y.
\end{split} 
\right. 
\end{equation}
\end{proposition}

This proposition can be proved by a direct calculation. 
Indeed, using (\ref{sdlinear}) and (\ref{sd_time2}),  we have 
\begin{align}
0 & = k \left( \psi_{n,y t_2} - \psi_{n,t_2 y} \right)
\nonumber \\
& = \left[ k q_{n,t_{2}} + q_{n,yy} + q_n \left( \gamma v_n + \delta v_{n+1} \right) + \gamma \delta \left( q_n r_n \right)^2 q_n \right]
	\left( \gamma \phi_n + \delta \phi_{n+1} \right)
\nonumber \\
& \hphantom{=} \; \, \mbox{}
	+ \gamma \delta q_n \left[ v_{n+1}-v_{n} + 2 \left( r_n q_n \right)_y \right] r_n \psi_n,  
\nonumber 
\end{align}
and 
\begin{align}
0 &= k \left( r_{n} \psi_{n} + \phi_{n} - \phi_{n+1} \right)_{t_2} 
\nonumber \\
& = \left[ k r_{n,t_{2}} - r_{n,yy} - \left( \delta v_n + \gamma v_{n+1} \right) r_n - \gamma \delta r_n \left( q_n r_n \right)^2 
	\right] \psi_{n}
\nonumber \\
& \hphantom{=} \; \, \mbox{}
	+ \left[ v_{n} -  v_{n+1} - 2 \left( r_n q_n \right)_y \right] \left( \gamma \phi_n + \delta \phi_{n+1} \right),
\nonumber 
\end{align}
which imply (\ref{second_flow}) for 
generic 
$\psi_n$ and $\phi_n$. 

By setting \mbox{$k=\mathrm{i}$} and \mbox{$\gamma=\delta=1 
$}
and imposing the 
Hermitian conjugation reduction 
\mbox{$r_n= - \varDelta q_n^\dagger$} 
and \mbox{$v_n^\dagger = v_n$}
on (\ref{second_flow}) 
where 
$\varDelta$ is a real-valued lattice parameter, 
we obtain 
\begin{equation} 
\label{reduced_second_flow}
\left\{ 
\begin{split}
& \mathrm{i} q_{n,t_{2}} + q_{n,yy} + q_n \left( v_n + v_{n+1} \right) + 
\varDelta^2 \sca{q_n}{q_n^\ast}^2
q_n =0, 
\\[2pt]
& v_{n+1}-v_{n} = 2 \varDelta \left( q_n^\dagger q_n \right)_y.
\end{split} 
\right. 
\end{equation}
Clearly, 
in the case where \mbox{$M=1$}, i.e., $q_n$ is a scalar, 
(\ref{reduced_second_flow}) provides an integrable semi-discretization 
of the elementary Davey--Stewartson flow (\ref{element2}), 
up to a rescaling of $q_n$.

\section{Integrable semi-discretizations of 
the 
Davey--Stewartson 
system 
and 
the 
\mbox{$(2+1)$}-dimensional Yajima--Oikawa system} 

We first establish the commutativity of the 
semi-discrete 
flow 
(\ref{first_flow}) and the semi-discrete 
flow
(\ref{second_flow}).  
Using (\ref{first_flow}) and (\ref{second_flow}), 
we have the relations: 
\begin{equation}
\mathrm{i} v_{n+1, t_1} - 2 \left( \alpha u_{n+1}^\gamma r_{n+1} q_n - \beta u_{n+1}^\delta r_n q_{n+1} \right)_y
= \mathrm{i} v_{n, t_1} - 2 \left( \alpha u_{n}^\gamma r_{n} q_{n-1} - \beta u_{n}^\delta r_{n-1} q_{n} \right)_y, 
\nonumber
\end{equation}
and 
\begin{align}
k \left( \log u_n \right)_{y t_2} &= \left[ q_{n,y} r_n - q_n r_{n,y} + \left( \gamma - \delta \right) \left( q_n r_n \right)^2 \right]_y 
\nonumber \\
& \hphantom{=} \; \, \mbox{ }- \left[ q_{n-1,y} r_{n-1} - q_{n-1} r_{n-1,y} + \left( \gamma - \delta \right) \left( q_{n-1} r_{n-1} 
	\right)^2 \right]_y, 
\nonumber
\end{align}
which imply 
\begin{align}
\mathrm{i} v_{n, t_1} &= 2 \left( \alpha u_{n}^\gamma r_{n} q_{n-1} - \beta u_{n}^\delta r_{n-1} q_{n} \right)_y + \cal{F} 
\nonumber \\
&= 2 \alpha \gamma u_{n}^\gamma \left( q_{n-1} r_{n-1} - q_{n} r_{n} \right) r_{n} q_{n-1}
	+ 2 \alpha u_{n}^\gamma \left( r_{n} q_{n-1} \right)_y 
\nonumber \\
& \hphantom{=} \; \, \mbox{ } - 2 \beta \delta u_{n}^\delta \left( q_{n-1} r_{n-1} - q_{n} r_{n} \right) r_{n-1} q_{n} 
	- 2 \beta u_{n}^\delta \left( r_{n-1} q_{n} \right)_y + \cal{F}, 
\label{v_t_1}
\end{align}
and 
\begin{align}
k \left( \log u_n \right)_{t_2} &= \left[ q_{n,y} r_n - q_n r_{n,y} + \left( \gamma - \delta \right) \left( q_n r_n \right)^2 \right]
\nonumber \\
& \hphantom{=} \; \, \mbox{ }- \left[ q_{n-1,y} r_{n-1} - q_{n-1} r_{n-1,y} + \left( \gamma - \delta \right) \left( q_{n-1} r_{n-1} 
	\right)^2 \right] + {\cal G}_n,
\label{log_u_n_t_2}
\end{align}
respectively, 
where $\cal{F}$ is an $n$-independent \mbox{$M \times M$} matrix and 
${\cal G}_n$ is a $y$-independent scalar. 
Using (\ref{first_flow}), 
(\ref{second_flow}) 
and (\ref{log_u_n_t_2}), 
we also obtain 
\begin{align}
k w_{n,y t_2} &= \left( \left\{ \beta \gamma u_{n}^\delta 
	\left[ -q_{n,y} r_{n-1} + q_n r_{n-1,y} + \delta \left( q_{n-1} r_{n-1} + q_n r_n \right) q_n r_{n-1} \right] \right. \right. 
\nonumber \\
& \hphantom{=} \; \, \left. \mbox{} + \alpha \delta u_{n}^\gamma 
	\left[ -q_{n-1,y} r_{n} + q_{n-1} r_{n,y} - \gamma \left( q_{n-1} r_{n-1} + q_n r_n \right) q_{n-1} r_{n} \right] \right\}_y
\nonumber \\
& \hphantom{=} \; \, \left. \mbox{} + \beta \gamma \delta u_{n}^\delta {\cal G}_n q_n r_{n-1} 
	+ \alpha \gamma \delta u_n^\gamma {\cal G}_n q_{n-1} r_n \right) - \left( n \to n+1 \right). 
\label{w_n_y_t_2}
\end{align}
Thus, we assume 
\mbox{${\cal G}_n=0$} to 
obtain 
\begin{align}
k \left( \log u_n \right)_{t_2} &= \left[ q_{n,y} r_n - q_n r_{n,y} + \left( \gamma - \delta \right) \left( q_n r_n \right)^2 \right]
\nonumber \\
& \hphantom{=} \; \, \mbox{ }- \left[ q_{n-1,y} r_{n-1} - q_{n-1} r_{n-1,y} + \left( \gamma - \delta \right) \left( q_{n-1} r_{n-1} 
	\right)^2 \right],
\label{log_u_n_t_2'}
\end{align}
and 
\begin{align}
k w_{n,t_2} &= \beta \gamma u_{n}^\delta 
	\left[ -q_{n,y} r_{n-1} + q_n r_{n-1,y} + \delta \left( q_{n-1} r_{n-1} + q_n r_n \right) q_n r_{n-1} \right] 
\nonumber \\
& \hphantom{=} \; \, \mbox{} - \beta \gamma u_{n+1}^\delta 
	\left[ -q_{n+1,y} r_{n} + q_{n+1} r_{n,y} + \delta \left( q_{n} r_{n} + q_{n+1} r_{n+1} \right) q_{n+1} r_{n} \right] 
\nonumber \\
& \hphantom{=} \; \, \mbox{} 
+ \alpha \delta u_{n}^\gamma 
	\left[ -q_{n-1,y} r_{n} + q_{n-1} r_{n,y} - \gamma \left( q_{n-1} r_{n-1} + q_n r_n \right) q_{n-1} r_{n} \right]
\nonumber \\
& \hphantom{=} \; \, \mbox{} -  \alpha \delta u_{n+1}^\gamma 
	\left[ -q_{n,y} r_{n+1} + q_{n} r_{n+1,y} - \gamma \left( q_{n} r_{n} + q_{n+1} r_{n+1} \right) q_{n} r_{n+1} \right] + {\cal H}_n, 
\label{w_n_y_t_2'}
\end{align}
where ${\cal H}_n$ is a $y$-independent scalar. 

By a direct calculation, we arrive at the following proposition. 
\begin{proposition}
\label{}
Equations 
\mbox{$(\ref{first_flow})$}, \mbox{$(\ref{second_flow})$}, \mbox{$(\ref{v_t_1})$}, 
\mbox{$(\ref{log_u_n_t_2'})$} 
and \mbox{$(\ref{w_n_y_t_2'})$}
imply 
that 
the two differentiation operators $\partial_{t_1}$ and $\partial_{t_2}$ commute, i.e., 
\begin{equation}
\nonumber 
q_{n, t_1 t_2} = q_{n, t_2 t_1} \;\, \mathrm{and} \;\, r_{n, t_1 t_2} = r_{n, t_2 t_1}, 
\end{equation}
if and only if  the ``constants'' of integration ${\cal F}$ and ${\cal H}_n$ 
satisfy the condition \mbox{$\left( \gamma + \delta \right) {\cal F} + {\cal H}_n I_M = 0$}, 
where $I_M$ is the \mbox{$M \times M$} identity matrix. 
Therefore, 
${\cal F}$ 
should be $y$-independent and ${\cal H}_n$ 
should be $n$-independent, namely, 
\mbox{${\cal F}_y=0$} and \mbox{${\cal H}_n={\cal H}$}; 
by 
the 
change of dependent variables 
\begin{equation}
\nonumber
q_n= \widetilde{q}_n \exp \left( -\frac{\mathrm{i}}{k} \int \int {\cal H} \, \mathrm{d} t_1 \mathrm{d} t_2 \right), 
\hspace{5mm}
r_n= \widetilde{r}_n \exp \left( \frac{\mathrm{i}}{k} \int \int {\cal H} \, \mathrm{d} t_1 \mathrm{d} t_2 \right), 
\end{equation}
we can set \mbox{${\cal F}=0$} and \mbox{${\cal H}=0$} without loss of generality. 
\end{proposition}

The commutativity of the 
semi-discrete 
flow 
(\ref{first_flow}) and the semi-discrete 
flow
(\ref{second_flow}) 
motivates us to consider a linear combination of these 
two flows: 
\begin{equation}
\nonumber 
\mathrm{i} \partial_t := \mathrm{i} \partial_{t_1} + 
bk \partial_{t_2}, 
\end{equation}
where $b$ is a nonzero constant. 
The corresponding time evolution of the linear wavefunction reads  
\begin{subnumcases}{\label{sd_time_g}}
\mathrm{i} \psi_{n,t} = \alpha u_n^\gamma \psi_{n-1} + \beta u_{n+1}^\delta \psi_{n+1} + w_n \psi_{n}
\nonumber \\ \hspace{15mm} \hspace{1pt}
	\mbox{} + b \left\{ \left( \gamma + \delta \right) q_n \phi_{n,y}
	- \left( \gamma + \delta \right) 
	\left( q_{n,y} - \delta q_n r_n q_n \right) \phi_n \right. 
\nonumber \\ \hspace{15mm} \left. 
	\mbox{} - \delta \left[ \left( q_{n,y} r_n - q_n r_{n,y} \right) 
	+ \left( \gamma - \delta \right) \left( q_n r_n \right)^2 \right] \psi_{n} \right\}, 
\label{sd_time_g1}
\\[2pt]
\mathrm{i} \phi_{n,t} =  -\alpha u_n^\gamma r_n \psi_{n-1}  + \beta u_{n}^\delta r_{n-1} \psi_{n}
	+ b \left[ \phi_{n,yy}  + \left( \gamma + \delta \right) v_{n} \phi_{n} \right],
\label{sd_time_g2}
\end{subnumcases}
where 
$\alpha$, $\beta$, $\gamma$ and $\delta$ are 
constants
satisfying the conditions \mbox{$\gamma+\delta \neq 0$} and 
\mbox{$(\alpha \gamma, \beta \delta) \neq (0,0)$}.

By a direct calculation, we can prove the following proposition. 

\begin{proposition}
\label{prop3.2}
The compatibility 
conditions of the 
overdetermined 
linear 
systems 
\mbox{$(\ref{sdlinear})$} and \mbox{$(\ref{sd_time_g})$} 
for $\psi_n$ and $\phi_n$ 
are equivalent to the system of 
differential-difference equations: 
\begin{equation} 
\label{first+second_flow}
\left\{ 
\begin{split}
& \mathrm{i} q_{n,t}  - \alpha u_n^\gamma q_{n-1} - \beta u_{n+1}^\delta q_{n+1} - w_n q_n 
\\ 
& \hphantom{\mathrm{i} q_{n,t}}
+ b \left[ q_{n,yy} + q_n \left( \gamma v_n + \delta v_{n+1} \right) + \gamma \delta \left( q_n r_n \right)^2 q_n \right] 
=0, 
\\[3pt]
& \mathrm{i} r_{n,t} + \beta u_{n}^\delta r_{n-1} + \alpha u_{n+1}^\gamma r_{n+1} +w_n r_n  
\\ & \hphantom{\mathrm{i} r_{n,t}}
- b \left[ r_{n,yy} + \left( \delta v_n + \gamma v_{n+1} \right) r_n + \gamma \delta r_n \left( q_n r_n \right)^2 \right] 
=0, 
\\[3pt]
& u_{n,y} = u_n \left( q_{n-1} r_{n-1} - q_n r_n \right), 
\\[3pt]
& w_{n,y} = \alpha \delta \left( u_n^{\gamma} q_{n-1} r_{n} 
	- u_{n+1}^{\gamma} q_n r_{n+1} \right) 
	+ \beta \gamma \left( u_n^{\delta} q_{n} r_{n-1} 
	- u_{n+1}^{\delta} q_{n+1} r_{n} \right), 
\\[3pt]
& v_{n+1} -v_{n} = - 2 \left( r_n q_n \right)_y.
\end{split} 
\right. 
\end{equation}

\end{proposition}

Under the parametric conditions
\begin{equation}
\beta=\alpha^\ast, \hspace{5mm} \gamma=\delta \in \mathbb{R}, 
\hspace{5mm} b \in \mathbb{R},
\nonumber
\end{equation}
the system (\ref{first+second_flow}) 
admits 
the Hermitian conjugation reduction: 
\begin{equation}
r_n =  - \varDelta q_n^\dagger, \hspace{5mm} u_n^\ast=u_n, \hspace{5mm} w_n^\ast = w_n, \hspace{5mm} v_n^\dagger = v_n, 
\nonumber
\end{equation}
where 
$\varDelta$ is 
a real-valued lattice parameter. 

In particular, 
if 
\mbox{$\alpha=\beta (=: -a) \in \mathbb{R}_{\neq 0}$} 
and \mbox{$\gamma=\delta=1$}, 
this reduction 
with a rescaling of $w_n$ as 
\mbox{$w_n =: -a \mathcal{W}_n 
$}
simplifies 
(\ref{first+second_flow}) to 
\begin{equation} 
\label{semi-discrete_DS}
\left\{ 
\begin{split}
& \mathrm{i} q_{n,t} 
+a \left[ u_{n+1} q_{n+1} + \mathcal{W}_n q_n + u_n q_{n-1} \right] 
\\ 
& \hphantom{\mathrm{i} q_{n,t}} + b \left[ q_{n,yy} + q_n \left( v_n + v_{n+1} \right) 
	+ \varDelta^2 \sca{q_n}{q_n^\ast}^2
q_n \right] =0, 
\\[2pt]
& u_{n,y} = \varDelta u_n \left( \sca{q_n}{q_n^\ast} 
- \sca{q_{n-1}}{q_{n-1}^\ast} \right) 
, 
\\[2pt]
& \mathcal{W}_{n,y} = \varDelta u_{n+1} \left( \sca{q_n}{q_{n+1}^\ast} + \sca{q_{n+1}}{q_n^\ast} \right)
 - \varDelta u_n \left( \sca{q_{n-1}}{q_n^\ast} + \sca{q_n}{q_{n-1}^\ast} \right),
\\[2pt]
& v_{n+1}-v_{n} = 2 \varDelta \left( q_n^\dagger q_n \right)_y. 
\end{split} 
\right. 
\end{equation}
Here, 
$a$ and $b$ are nonzero real constants, 
\mbox{$q_n \in \mathbb{C}^M$}, \mbox{$u_n, \mathcal{W}_n \in \mathbb{R}$} 
and $v_n$ is an \mbox{$M \times M$} Hermitian matrix. 

In the case where \mbox{$M=1$}, i.e., $q_n$ is a scalar, 
(\ref{semi-discrete_DS}) provides an integrable semi-discretization 
of the Davey--Stewartson system (\ref{continuousDS}), 
up to a rescaling of $q_n$. 
Indeed, 
by setting 
\begin{align}
\nonumber
& q_n = q(n \Delta, y, t), \hspace{5mm} u_n = \frac{1}{\Delta^2} + \frac{1}{2} u(n \Delta, y, t), 
\\[2mm]
& \nonumber 
\mathcal{W}_n = -\frac{2}{\Delta^2} + \mathcal{W} (n \Delta, y, t), \hspace{5mm}
v_n = v (n \Delta, y, t),
\end{align}
and taking the continuous limit \mbox{$\Delta \to 0$}, 
(\ref{semi-discrete_DS})
with 
scalar $q_n$
reduces to
\begin{equation} 
\nonumber
\left\{ 
\begin{split}
& \mathrm{i} q_{t} 
+a \left[ q_{xx} + \left( u + \mathcal{W} \right) q \right] + b \left( q_{yy} + 2 q v \right) =0, 
\\[2pt]
& u_{y} = \mathcal{W}_{y} = 2 \left( \left| q \right|^2 \right)_x, 
\\[2pt]
& v_{x} = 2 \left( \left| q \right|^2 \right)_y,
\end{split} 
\right. 
\end{equation}
where \mbox{$x:= n \Delta$}. 

By applying the linear change of the independent variables 
\begin{equation}
\widetilde{t} = t_1 + y, \hspace{5mm} \widetilde{y}= c y, 
\label{Galilean-like2}
\end{equation}
where $c$ is an arbitrary real constant, 
to 
the semi-discrete elementary Davey--Stewartson flow (\ref{reduced_first_flow2}), 
we 
obtain 
\begin{equation} 
\label{sd_YO1}
\left\{ 
\begin{split}
& \mathrm{i} q_{n,t} 
= u_{n+1} q_{n+1} + w_n q_n + u_n q_{n-1} 
, 
\\[2pt]
& u_{n,t}+c u_{n,y} = \varDelta u_n \left( \sca{q_n}{q_n^\ast} 
- \sca{q_{n-1}}{q_{n-1}^\ast} \right) 
, 
\\[2pt]
& w_{n,t}+c w_{n,y} = \varDelta u_{n+1} \left( \sca{q_n}{q_{n+1}^\ast} + \sca{q_{n+1}}{q_n^\ast} \right)
 - \varDelta u_n \left( \sca{q_{n-1}}{q_n^\ast} + \sca{q_n}{q_{n-1}^\ast} \right),
\end{split} 
\right. 
\end{equation}
where 
the tilde of the continuous independent variables 
is omitted for notational brevity. 
The system (\ref{sd_YO1}) 
with scalar 
\mbox{$q_n \in \mathbb{C}$} 
and 
\mbox{$u_n, w_n \in \mathbb{R}$} 
provides an integrable semi-discretization of 
the \mbox{$(2+1)$}-dimensional Yajima--Oikawa system (\ref{2DYO}), 
up to a rescaling of variables; 
in the absence of $y$-dependence, 
it reduces to 
the discrete 
Yajima--Oikawa system 
proposed in our previous paper~\cite{Tsuchida18-2}. 
We remark that 
another integrable semi-discretization of 
the \mbox{$(2+1)$}-dimensional Yajima--Oikawa system (\ref{2DYO}) 
was proposed 
recently in~\cite{Yu15}. 

\section{Concluding remarks}

In this paper, we 
discussed 
the problem of 
how to discretize one of the two spatial variables 
in the Davey--Stewartson system (\ref{continuousDS}) 
and the \mbox{$(2+1)$}-dimensional Yajima--Oikawa system (\ref{2DYO}). 
To preserve the integrability, 
we considered the linear problem (\ref{sdlinear}); 
by 
assuming 
\mbox{$\gamma=\delta \in \mathbb{R}$}
and imposing 
the Hermitian conjugation reduction 
\mbox{$r_n =  - \varDelta q_n^\dagger$} 
where 
$\varDelta$ is 
a real constant,  
(\ref{sdlinear}) can be understood 
as a semi-discrete analog of (the vector generalization of) 
the continuous linear problem (\ref{clinear_s}). 
By choosing the associated time evolution of the linear wavefunction appropriately, 
we obtain
(\ref{reduced_first_flow2}) 
(resp.~(\ref{reduced_second_flow})) 
as 
(a vector generalization of)
an integrable semi-discretization 
of the elementary Davey--Stewartson flow (\ref{element1}) (resp.~(\ref{element2})). 
The two flows (\ref{reduced_first_flow2}) and (\ref{reduced_second_flow}) 
commute under a suitable choice of the ``constants'' of integration, 
so we can naturally consider a linear combination of them 
to obtain (\ref{semi-discrete_DS}) 
as (a vector generalization of) an integrable semi-discretization 
of the Davey--Stewartson system (\ref{continuousDS}). 

By applying the linear change of the independent variables (\ref{Galilean-like2}) 
to 
the 
flow 
(\ref{reduced_first_flow2}) 
and omitting the tilde of the continuous independent variables, 
we obtain the system (\ref{sd_YO1}); 
this system provides 
(a vector generalization of) an integrable semi-discretization of 
the \mbox{$(2+1)$}-dimensional Yajima--Oikawa system (\ref{2DYO}),  
%
%
%
which is 
a \mbox{$(2+1)$}-dimensional generalization of 
the 
discrete Yajima--Oikawa system
proposed 
in our previous paper~\cite{Tsuchida18-2}.


\end{document}